\documentclass[%
 aip,
 jcp,%
 amsmath,amssymb,
reprint,%
]{revtex4-1}

\usepackage{graphicx}
\usepackage{dcolumn}
\usepackage{bm}
\usepackage{color}

\begin{document}



\title[]{Beyond Marcus theory and the Landauer-B\"{u}ttiker approach in molecular junctions. II. A self-consistent Born approach}

\author{Jakub K. Sowa}
 \email{jakub.sowa@northwestern.edu}
\affiliation{ 
Department of Chemistry, Northwestern University, Evanston, IL 60208, USA}%
\affiliation{ 
Department of Materials, University of Oxford, OX1 3PH Parks Road, Oxford, UK}%
\author{Neill Lambert}
\affiliation{ 
Theoretical Quantum Physics Laboratory, RIKEN Cluster for Pioneering Research, Wako-shi, Saitama 351-0198, Japan}%
\author{Tamar Seideman}
\affiliation{ 
Department of Chemistry, Northwestern University, Evanston, IL 60208, USA}%
\author{Erik M. Gauger}
\affiliation{ 
SUPA, Institute of Photonics and Quantum Sciences, Heriot-Watt University, Edinburgh EH14 4AS, UK}%

\date{\today}

\begin{abstract} 
Marcus and Landauer-B\"uttiker approaches to charge transport through molecular junctions describe two contrasting mechanisms of electronic conduction.
In previous work, we have shown how these charge transport theories can be unified in the single-level case by incorporating lifetime broadening into the second-order quantum master equation. Here, we extend our previous treatment by incorporating lifetime broadening in the spirit of the self-consistent Born approximation. By comparing both theories to numerically converged hierarchical-equations-of-motion (HEOM) results, we demonstrate that our novel self-consistent approach rectifies shortcomings of our earlier framework which are present especially in the case of relatively strong electron-vibrational coupling.
We also discuss circumstances under which the theory developed here simplifies to the generalised theory developed in our earlier work.  
Finally, by considering the high-temperature limit of our new self-consistent treatment, we show how lifetime broadening can also be self-consistently incorporated into Marcus theory.
Overall, we demonstrate that the self-consistent approach constitutes a more accurate description of molecular conduction while retaining most of the conceptual simplicity of our earlier framework.
\end{abstract}

\maketitle

\section{\label{Intro}Introduction}
The vast majority of experimental studies of molecular conduction has been performed in the off-resonant transport regime, that is when the molecular energy levels lie outside the bias window.
This regime is nowadays very well understood within the framework of the  Landauer-B\"uttiker approach.\cite{landauer1957spatial,nitzan2006chemical,nitzan2001electron} Typically, the transmission function, placed at the heart of that approach, is obtained using non-equilibrium Green's function techniques coupled with density functional theory,\cite{xue2002first} often yielding a good match between the predicted and observed behaviour.\cite{lindsay2007molecular} 
Over the last few years, however, there has been a growing experimental interest in the resonant transport regime.\cite{fung2019breaking,thomas2019understanding,gelbwaser2018high,du2018terahertz} This regime can be achieved by applying appropriately high bias voltage\cite{zang2018resonant} or by electrostatically gating the molecular structure within the junction.\cite{perrin2016gate,perrin2015single,burzuri2016sequential,lau2015redox,gehring2017distinguishing,limburg2019charge} 
It is well-known that electron-vibrational (and electron-electron) interactions cannot be ignored in this scenario,\cite{thomas2019understanding} in contrast to what is often justifiable in the off-resonant regime.\cite{galperin2007molecular,gehring2019single}
Consequently, the resonant transport regime often delivers a much richer breadth of phenomena\cite{koch2005franck,hartle2011resonant,sowa2017vibrational,volkovich2011bias,hartle2011vibrational} but also constitutes a more formidable theoretical problem.

A number of theoretical approaches to describe vibrational effects in the resonant transport regime of molecular junctions have been developed.\cite{galperin2007molecular,jorn2009competition,jorn2006theory,simine2013path,wang2011numerically,muhlbacher2008real,mitra2004phonon,jorn2010implications,souto2014dressed} Several of them deserve a special mention, including the non-equilibrium Green's function (NEGF) approach of Ref.~\onlinecite{galperin2006resonant}~(building on earlier NEGF efforts\cite{flensberg2003tunneling,lundin2002temperature,jauho1994time,wingreen1989inelastic}) as well as numerically exact hierarchical equation of motion (HEOM) methods.\cite{erpenbeck2018extending,schinabeck2016hierarchical,schinabeck2018hierarchical,jiang2012inelastic}
Because of their simplicity, the (second-order) rate-equation and quantum-master-equation techniques also constitute a popular theoretical framework for investigating the phenomena in question.\cite{koch2005franck,bruus2004many,peskin2019quantum,sowa2017environment,seldenthuis2008vibrational,santamore2013vibrationally}
Due to their perturbative nature, however, such approaches are typically valid only at relatively high temperatures (or for very weak molecule-lead coupling) in the resonant transport regime.\footnote{Quantum master equation approaches are also useful (and accurate) in the case of very high bias especially if the intramolecular dynamics plays an important role in the overall charge transport.\cite{gurvitz1996microscopic}} Their applicability can be extended, however, by incorporating lifetime broadening into these theoretical frameworks.
This can be done, for instance, by also considering higher-order terms in the perturbative expansion\cite{koch2006theory,bruus2004many,schinabeck2016hierarchical} or by replacing the (free) system propagator (as present within the perturbative theory) with a dressed one. \cite{esposito2009transport,esposito2010self,jin2014improved,sowa2018beyond}

In earlier work (Ref.~\onlinecite{sowa2018beyond}, henceforth referred to as Part I), some of us have studied a spin-less Anderson-Holstein model within the latter approach.
We have shown how lifetime broadening can be introduced into the second-order master equation by replacing the free propagator with an effective one which was obtained by considering the equations of motion for the relevant fermionic operators. Recently, it has also been shown that the same result can be obtained using a generalised input-output approach.\cite{liu2019generalized}
As we have discussed in our work, however, introducing lifetime broadening in such a way can lead to an overestimation of this effect, especially in the presence of strong electron-vibrational coupling to low-frequency vibrational modes.  
In order to improve upon our earlier work  we here demonstrate how the effective propagator can be obtained in the spirit of the self-consistent Born approximation,\cite{cui2006quantum} in analogy to what has been previously done in the case of purely electronic quantum transport.\cite{esposito2010self,jin2014improved,li2016number}
The superiority of this (non-iterative) self-consistent method (as compared to our earlier efforts) will be demonstrated by comparing both approaches to HEOM calculations.

We organise this work as follows:
First, in Section \ref{model}, we discuss the model of the molecular junction used in this work. In Section \ref{GQ} we next derive a self-consistent expression for the electric current and, in Section \ref{GQ2}, discuss its limitations and demonstrate its effectiveness in the case of coupling to a single vibrational mode.
In Section \ref{highT}, we consider the high-temperature limit of our self-consistent approach. This yields a Marcus-type theory of transport similar to that derived in Part I, but one in which lifetime broadening is introduced in a self-consistent fashion. 
We conclude with a brief summary in Section \ref{end}.

\section{Theoretical Approach}
\subsection{Model \label{model}}
\begin{figure}
    \centering
    \includegraphics{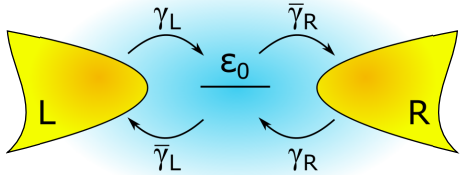}
    \caption{Schematic illustration of the system considered in this work. The molecular system comprises a single electronic level which is coupled to two  [left (L) and right (R)] metallic electrodes as well as a wider phononic environment. $\gamma_l$ and $\bar{\gamma}_l$ denote rates of electron hopping on and off the molecular system, respectively, see below.}
    \label{scheme}
\end{figure}

The theoretical model used in this work is identical to the one used by us in Part I, and schematically shown in Fig.~\ref{scheme}. Nonetheless, for completeness and convenience, we briefly describe it below. The molecular junction is governed by the following Hamiltonian:
\begin{equation}
    H = H_S + H_{SE} + H_E + H_{SB} + H_{B}~.
\end{equation}

The molecular system is modelled as a single electronic energy level with energy $\varepsilon_0$  and the creation (annihilation) operator $a_0^\dagger$ ($a_0$):
\begin{equation}
    H_S = \varepsilon_0 \: a_0^\dagger a_0 ~.
\end{equation}
The position of this energy level can be altered by applying the gate potential $V_\mathrm{g}$: $\varepsilon_0 = \varepsilon_{00} - \lvert e\rvert V_\mathrm{g}$ where $e$ is the electron charge.
This molecular level is coupled to a left (L) and a right (R) fermionic reservoir (lead) which are governed by 
\begin{equation}
    H_E = \sum_{l=\mathrm{L},\mathrm{R}} \sum_{k_l} \epsilon_{k_l} c^\dagger_{k_l} c_{k_l} ~,
\end{equation}
where $c^\dagger_{k_l}$ ($c_{k_l}$) is the creation (annihilation) operator for an electron in the level $k_l$ with energy $\epsilon_{k_l}$ in the lead $l$.
The coupling between the molecule and the reservoirs is accounted for by
\begin{equation}
    H_{SE} = \sum_{l=\mathrm{L},\mathrm{R}} \sum_{k_l} V_{k_l} a^\dagger_0  c_{k_l}  + V_{k_l}^*c^\dagger_{k_l} a_0 ~,
\end{equation}
where $V_{k_l}$ is the electronic coupling strength. 
The molecular energy level also interacts with its vibrational  (phononic) environment modelled as a collection of harmonic oscillators with frequencies $\omega_q$, and lowering  and raising operators $b_q$ and $b_q^\dagger$, respectively:  
\begin{equation}
  H_{B} = \sum_{q} \: \omega_q b_q^\dagger b_q  ~.
\end{equation}
The electronic-vibrational interactions Hamiltonian is given by
\begin{equation}
  H_{SB} = \sum_{q} g_q\: a_0^\dagger a_0 (b_q^\dagger + b_q)~,
\end{equation}
where $g_q$ is the coupling constant for the interaction between the molecular energy level and the vibrational mode $q$.

Throughout this work, we will assume that the vibrational environment is thermalised at all times. This constitutes an important approximation which is more likely to be valid for the broader (outer-sphere) background environment and in the limit of relatively weak molecule-electrode coupling (when the time between individual electron transfers between the leads and the molecular system is longer than the environmental relaxation time),\cite{migliore2011nonlinear} see also the discussion below.

\subsection{Numerically converged benchmark}

For the numerically exact HEOM calculations we follow closely the treatment used by Schinabeck {\em et al.}\cite{schinabeck2016hierarchical,jiang2012inelastic} As mentioned, we assume a thermalised vibrational environment by absorbing the displacement operators $X$ and $X^{\dagger}$ [see Eq.~(\ref{eq:displacements}) below] into the lead-coupling, and implicitly truncating the resulting correlation functions at second order (see Section 2.7.5 of Ref.~\onlinecite{schinabeckDissertation} for a detailed discussion).  In other words we treat the electronic leads exactly, but impose an approximation on the vibrational environment that is consistent with the ones we use the in other approaches in this work.  For the leads we use the Pad\'{e} decomposition of the correlation functions, which is truncated at value $L_{\mathrm{max}}$ when convergence of the electric current is observed.  We must also truncate the number of exponents that arise in the expansion of the vibrational environment correlation function [see Eq.~\eqref{a3} in the Appendix \ref{appFou}], wherein the truncation is performed on the indices of the summation at some value $u_{\mathrm{max}}$ which gives convergence in the observed results.

\section{Self-consistent Theory: Derivation\label{GQ}}
The first step in our derivation of the self-consistent master equation is identical to that presented in Part I. We begin with the Lang-Firsov transformation\cite{lang1963kinetic,mahan2013many} which is given by
\begin{equation}
\bar{H} = e^G H e^{-G}~,
\end{equation}
where $G =\sum_{q} (g_{q}/\omega_q)\: a^\dagger_0 a_0(b^\dagger_{q} - b_{q})$. 
This transformation removes the electron-phonon coupling term from the Hamiltonian, renormalises the energy of the molecular level, and introduces the displacement operators  $X$ and $X^\dagger$ into the terms describing the molecule-lead coupling. The polaron-transformed Hamiltonian takes the form:
\begin{align}
    &\bar{H} = \bar{H}_S + \bar{H}_{SE} + H_E + H_{B}~; \label{polaron}\\
    &\bar{H}_S =  \bar{\varepsilon}_0 \: a_0^\dagger a_0 ~;\\
    &\bar{H}_{SE} = \sum_{l,k_l} V_{k_l} X^\dagger a^\dagger_0  c_{k_l}  + V_{k_l}^*c^\dagger_{k_l} X a_0  ~, \label{HSE}
\end{align}
where $\bar{\varepsilon}_0 = \varepsilon_0 - \sum_q \lvert g_q\rvert^2/\omega_q$ is the renormalised position of the molecular level, and (for real $g_q$) the displacement operators are given  by
\begin{equation}
X^\dagger = \exp\left[\sum_q \dfrac{g_q}{\omega_{q}}(b^\dagger_{q} - b_{q}) \right]~,
\label{eq:displacements}
\end{equation}
and equivalently for $X$. 

\subsection{Born-Markov quantum master equation}
We first consider the well-known second-order quantum master equation within the Born-Markov approximation ($\hbar = 1$ throughout):\cite{breuer2002theory}
\begin{equation} \label{Born}
    \dfrac{\mathrm{d}\rho(t)}{\mathrm{d}t} = -\mathrm{i}[\bar{H}_S, \rho(t)] - \int_0 ^\infty \mathrm{d}\tau \: \mathrm{Tr}[\bar{H}_{SE},[\bar{H}_{SE}(\tau),\rho(t)]] ~.
\end{equation}
In the above,  $\mathrm{Tr[...]}$ denotes a trace over all the environmental degrees of freedom, and $\bar{H}_{SE}(\tau)= e^{-\mathrm{i}H_0 \tau}\bar{H}_{SE}\: e^{\mathrm{i}H_0 \tau}$ where $H_0 = \bar{H}_S + H_E + H_B$.

We first simply expand the commutators present in Eq.~\eqref{Born} yielding:
\begin{multline} \label{expan}
\dfrac{\mathrm{d}\rho(t)}{\mathrm{d}t} = -\mathrm{i}[\bar{H}_S,\rho(t)] -  \sum_l\sum_{k_l}\int_0 ^\infty \mathrm{d}\tau \bigg\{ \bigg.\\
   C^+_{k_l}(\tau) B(\tau) a_0 \: \mathcal{G}(\tau) \left[ a_0^\dagger\right] \rho(t)
- {C^-_{k_l}}^*(\tau) B^*(\tau) a_0 \: \rho(t) \mathcal{G}(\tau) \left[a_0^\dagger\right] \\
+  C^-_{k_l}(\tau) B(\tau) a_0^\dagger \: \mathcal{G}(\tau) \left[ a_0\right] \rho(t)
- {C^+_{k_l}}^*(\tau) B^*(\tau) a_0^\dagger \: \rho(t) \mathcal{G}(\tau) \left[a_0\right] \\
+  {C^-_{k_l}}^*(\tau)B^*(\tau) \:\rho(t)  \mathcal{G}(\tau) \left[a_0^\dagger\right] a_0 
- C^+_{k_l}(\tau) B(\tau) \: \mathcal{G}(\tau) \left[ a_0^\dagger\right] \rho(t)a_0  \\
+  {C^+_{k_l}}^*(\tau) B^*(\tau) \:\rho(t) \mathcal{G}(\tau) \left[a_0\right]a_0^\dagger
- C^-_{k_l}(\tau) B(\tau) \: \mathcal{G}(\tau) \left[a_0 \right]\rho(t) a_0^\dagger \bigg.\bigg\}~.
\end{multline}
In the above, we have defined the free (system) propagators:
\begin{equation}
\mathcal{G}(\tau)[A] =e^{-\mathrm{i}\bar{H}_S \tau}A\: e^{\mathrm{i}\bar{H}_S \tau}~.
\end{equation}
The fermionic correlation functions in Eq.~\eqref{expan} are:
$ C^+_{k_l}(\tau) = \lvert V_{k_l}\rvert^2 \langle c^\dagger_{k_l}(\tau) c_{k_{l}} \rangle$ and $C^-_{k_l}(\tau) = \lvert V_{k_l}\rvert^2 \langle c_{k_l}(\tau) c_{k_{l}}^\dagger \rangle$.
Assuming that the metallic leads possess a continuum of energy levels, we express them as:
\begin{equation}\label{fourier}
    \sum_{k_l} C^\pm_{k_l}(t)  = \int_{-\infty}^\infty \dfrac{\mathrm{d}\epsilon}{2\pi} \ \Gamma_l^\pm (\epsilon) \ e^{\pm\mathrm{i}\epsilon t}~,
\end{equation}
where $\Gamma_l^+ (\epsilon) =  \Gamma_l(\epsilon) f_l(\epsilon)$ and
$\Gamma_l^- (\epsilon) =  \Gamma_l(\epsilon) [1 - f_l(\epsilon)]$.   
$\Gamma_l(\epsilon) = 2\pi \sum_{k_l} \lvert V_{k_l}\rvert^2 \delta(\epsilon- \epsilon_{k_l}) $ is the spectral density for the lead $l$, and $f_l(\epsilon)$ is the Fermi distributions in the lead $l$: $f_l(\epsilon) = (\exp[(\epsilon-\mu_l)/k_{\mathrm{B}} T] +1)^{-1}$, where $\mu_l$ is the corresponding chemical potential. In what follows, we shall take the so-called wide-band limit, i.e.~assume a constant density of states in the leads so that $\Gamma_l(\epsilon) = \Gamma_l = \mathrm{const.}$

Likewise, $B(\tau)$ in Eq.~\eqref{expan} is the phononic correlation function:
\begin{equation}
    B(\tau) = \langle X(\tau) X^\dagger\rangle~.
\end{equation}
As previously stated, throughout this work, we shall assume that the vibrational modes are in their thermal equilibrium state at all times. Therefore, the phononic correlation function $B(t)$ can be written as:\cite{mahan2013many}
\begin{multline}\label{corr}
    B(t) = \exp\bigg[\int_0^\infty \mathrm{d}\omega \dfrac{\mathcal{J}(\omega)}{\omega^2} \bigg( \coth{\left(\dfrac{\beta\omega}{2}\right)}\\ \times \big(\cos{\omega t} - 1\big) - \mathrm{i}\sin{\omega t} \bigg)\bigg] ~,
\end{multline}
where $\beta$ is the inverse temperature, $\beta = 1/k_{\mathrm{B}} T$, and where we have made use of the following definition of the phononic spectral density
\begin{equation}
    \mathcal{J}(\omega) = \sum_q \lvert g_q\rvert^2 \delta(\omega - \omega_q)
\end{equation}
which describes the distribution of the vibrational modes weighted by the strength of the electron-vibrational coupling.\\

The effect of the free propagator $\mathcal{G}(t)$ on the relevant operators can easily be found:
\begin{align}\label{g1}
   \mathcal{G}(t)[a_0] &= a_0 e^{\mathrm{i} \bar{\varepsilon}_0 t} ~;\\ 
   \mathcal{G}(t)[a_0^\dagger] &= a_0^\dagger e^{-\mathrm{i} \bar{\varepsilon}_0 t}~. \label{g2}
\end{align}
This yields the second-order (Born-Markov) quantum master equation:
\begin{multline} \label{BMQ}
\dfrac{\mathrm{d}\bar{\rho}(t)}{\mathrm{d}t} = -\mathrm{i}[\bar{H}_S,\rho(t)] + \sum_l \bigg\{ \upsilon_l \left(a_0^\dagger \rho(t) a_0 - a_0 a_0^\dagger \rho(t)\right) \\
+ \upsilon^*_l \left(a_0^\dagger \rho(t) a_0 - \rho(t) a_0 a_0^\dagger \right) 
+ \bar{\upsilon}_l \left(a_0 \rho(t) a_0^\dagger -  a_0^\dagger a_0 \rho(t)\right) \\
+ \bar{\upsilon}^*_l \left(a_0 \rho(t) a_0^\dagger - \rho(t) a_0^\dagger a_0  \right)  \bigg\}~,
\end{multline}
where the response functions $\upsilon_l$ and $\bar{\upsilon}_l$ are given by:
\begin{align}\label{rr1}
& \upsilon_l =   \Gamma_l \int_{-\infty}^\infty \dfrac{\mathrm{d}\epsilon}{2\pi} f_l(\epsilon)  \int_0^\infty \mathrm{d}\tau \ e^{+\mathrm{i}(\epsilon - \bar{\varepsilon}_0 )\tau}  B(\tau) ~;\\     
& \bar{\upsilon}_l =  \Gamma_l\int_{-\infty}^\infty \dfrac{\mathrm{d}\epsilon}{2\pi} [1 - f_l(\epsilon)]  \int_0^\infty \mathrm{d}\tau \ e^{-\mathrm{i}(\epsilon - \bar{\varepsilon}_0 )\tau} B(\tau) ~. \label{rr2}
\end{align}
For convenience, in Appendix \ref{appFou} we discuss how the Fourier transforms of the phononic correlation functions in Eqs.~\eqref{rr1} and \eqref{rr2} can be evaluated.

\subsection{Self-consistent quantum master equation}
Similarly to what we have done in the Part I of this work, in order to go beyond the second-order treatment, we go back to Eq.~\eqref{expan} and replace the free propagator $\mathcal{G}(t)$ with an effective one, $\mathcal{U}(t)$.
Replacing $\mathcal{G}(t)$ with an effective (interacting) propagator constitutes the crucial ansatz of our derivation.
Since this effective propagator will account for the molecule-lead  (as well as electron-vibrational) coupling,  it will introduce lifetime broadening into our theoretical description and, in principle, also account for the energy shift of the molecular level. Both of these effects are ignored within the conventional Born-Markov treatment.
In contrast to our earlier work however, here, we shall obtain $\mathcal{U}(t)$ in the spirit of the self-consistent Born approximation.\cite{cui2006quantum,jin2014improved}  That is, to find the effective propagator $\mathcal{U}(t)$ we will make use of the second-order Born-Markov quantum master equation [Eq.~\eqref{BMQ}] which itself contains the free propagator $\mathcal{G}(t)$. 

We define $\mathcal{U}(t)[a_0] \equiv a_0(t)$ and $\mathcal{U}(t)[a_0^\dagger] \equiv a^\dagger_0(t)$ where
\begin{multline}
  \dot{a}_0(t) = -\mathrm{i}[\bar{H}_S,a_0(t)] + \sum_l \bigg\{ \upsilon_l \left(a_0^\dagger a_0(t) a_0 - a_0 a_0^\dagger a_0(t)\right) \\
+ \upsilon^*_l \left(a_0^\dagger a_0(t) a_0 - a_0(t) a_0 a_0^\dagger \right) 
+ \bar{\upsilon}_l \left(a_0 a_0(t) a_0^\dagger -  a_0^\dagger a_0 a_0(t)\right) \\
+ \bar{\upsilon}^*_l \left(a_0 a_0(t) a_0^\dagger - a_0(t) a_0^\dagger a_0  \right)  \bigg\}~,  
\end{multline}
and similarly for $a_0^\dagger(t)$.
This master equation can be readily solved in the Liouville space yielding:
\begin{equation}
  \mathcal{U}(t)[a_0] = a_0(t) = \exp\left( {\mathrm{i} \bar{\varepsilon}_0 t} - \sum_l (\upsilon_l + \bar{\upsilon}^*_l) t\right) \: a_0~.
\end{equation}
Analogously, for $a_0^\dagger(t)$ we obtain:
\begin{equation}
  \mathcal{U}(t)[a_0^\dagger] = a_0^\dagger(t) = \exp\left( -{\mathrm{i} \bar{\varepsilon}_0 t} - \sum_l (\upsilon_l^* + \bar{\upsilon}_l) t\right) \: a_0^\dagger~.
\end{equation}
Unlike the effective propagator in Part I of this work, here the effect of $\mathcal{U}(t)$ depends on the value of the bias voltage and the strength of the electron-vibrational interactions, \textit{c.f.} Refs.~\onlinecite{flensberg2003tunneling,galperin2006resonant}.

The remaining steps in the derivation are relatively straightforward. 
We  insert the effective propagator analogously to what we have done within the Born-Markov approximation.
This self-consistent quantum master equation [which has the form identical to that of Eq.~\eqref{BMQ}] is then solved in the steady-state limit yielding the stationary density matrix:
\begin{equation}\label{SSS}
    \rho_{\mathrm{st}} = \Bigg(\begin{array}{c c}
  \dfrac{\bar{\gamma}_L +\bar{\gamma}_R}{\gamma_\mathrm{L} + \gamma_\mathrm{R} + \bar{\gamma}_L +\bar{\gamma}_R} & 0 \\ 
  0 &   \dfrac{\gamma_\mathrm{L} + \gamma_\mathrm{R}}{\gamma_\mathrm{L} + \gamma_\mathrm{R} + \bar{\gamma}_L +\bar{\gamma}_R}
 \end{array} \Bigg) ~,
\end{equation}
in the basis of two relevant charge states and where the rates in the above are:
\begin{align}\label{R1}
& \gamma_l =  2\ \Gamma_l  \mathrm{Re}  \int_{-\infty}^\infty \dfrac{\mathrm{d}\epsilon}{2\pi} f_l(\epsilon)  \int_0^\infty \mathrm{d}\tau \ e^{+\mathrm{i}(\epsilon - \bar{\varepsilon}_0 )\tau} e^{-\Phi \tau} B(\tau) ~;\\     
& \bar{\gamma}_l = 2\ \Gamma_l \mathrm{Re} \int_{-\infty}^\infty \dfrac{\mathrm{d}\epsilon}{2\pi} [1 - f_l(\epsilon)]  \int_0^\infty \mathrm{d}\tau \ e^{-\mathrm{i}(\epsilon - \bar{\varepsilon}_0 )\tau} e^{-\Phi^* \tau} B(\tau) ~. \label{R2}
\end{align}\\
The factor $\Phi$ in Eqs.~\eqref{R1} and \eqref{R2} is given by:
\begin{equation}
    \Phi = \sum_l (\upsilon_l + \bar{\upsilon}^*_l)~.
\end{equation}
The real part of $\Phi$ determines the amount of lifetime broadening  while the imaginary part of $\Phi$ induces renormalisations of the position of the molecular energy level (both of which will generally be bias-dependent).\\
Finally, we determine the stationary current flowing through the junction. It can be evaluated at either molecule-lead interface (in the steady-state limit the currents flowing through the left and the right interface are equal in magnitude) as $I = e \: [\gamma_l \rho_{00, \mathrm{st}} - \bar{\gamma}_l \rho_{11, \mathrm{st}}]$.\cite{flindt2004full}
The electric current is therefore, as in the Part I of our work, given by the generic expression:\cite{zhang2008single}
\begin{equation} \label{current}
    I = e \ \dfrac{\gamma_\mathrm{L} \bar{\gamma}_R - \gamma_\mathrm{R} \bar{\gamma}_L}{\gamma_\mathrm{L} + \gamma_\mathrm{R} + \bar{\gamma}_L +\bar{\gamma}_R} ~.
\end{equation}
The only difference between the self-consistent approach developed here and the generalised approach from Part I of this work lies therefore in the factor $\Phi$ [which in Part I was instead given by $\Gamma = (\Gamma_\mathrm{L} + \Gamma_\mathrm{R})/2$]. As we shall demonstrate, however, this seemingly small difference can  lead to large discrepancies between the behaviour predicted by these two approaches.

\section{Self-consistent Theory: Discussion \label{GQ2}}
\subsection{No vibrational coupling \label{Land}}
Let us begin by considering the limit of no vibrational coupling: $\mathcal{J}(\omega) = 0$. Then, the phononic correlation function becomes: $B(t) = 1$, and therefore
\begin{multline}
    \Phi = \sum_l  \Gamma_l \int_{-\infty}^\infty \dfrac{\mathrm{d}\epsilon}{2\pi} \int_0^\infty    \mathrm{d}\tau \ e^{+\mathrm{i}(\epsilon - \varepsilon_0 )\tau}= \\  = \sum_l  \Gamma_l \left[ \int_{-\infty}^\infty \dfrac{\mathrm{d}\epsilon}{2\pi} \pi \delta(\epsilon- \varepsilon_0) +  \mathcal{P} \int_{-\infty}^\infty \dfrac{\mathrm{d}\epsilon}{2\pi} \dfrac{\mathrm{i}}{\epsilon-\varepsilon_0} \right]\\ =\sum_l \Gamma_l/2~,
\end{multline}
where $\mathcal{P}$ denotes the Cauchy principal value.
As expected,\cite{reuter2009molecular} in the wide-band limit considered here, the renormalisation of the molecular energy level (imaginary part of $\Phi$) vanishes.
The hopping rates are then given by
\begin{equation}
    \gamma_l = 2\ \mathrm{Re}\ \Gamma_l \int_{-\infty}^\infty \dfrac{\mathrm{d}\epsilon}{2\pi} f_l(\epsilon) \int_0^\infty \mathrm{d}\tau e^{\mathrm{i}(\epsilon-\varepsilon_0)\tau} e^{-\Gamma \tau}~,
\end{equation}
and analogously for $\bar{\gamma}_l$, where we have again defined $\Gamma = \sum_l\Gamma_l/2$. 
As we have shown in Appendix A of Part I, the overall expression for the electric current then becomes
\begin{equation} \label{landauer}
    I = e \int_{-\infty}^\infty \dfrac{\mathrm{d}\epsilon}{2\pi} [f_L(\epsilon) - f_R(\epsilon)] \dfrac{\Gamma_\mathrm{L}\Gamma_\mathrm{R}}{(\epsilon - \varepsilon_0)^2  + \Gamma^2} ~.
\end{equation}
In the absence of electron-vibrational interactions, therefore, we again recover the conventional Landauer-B\"{u}ttiker result for charge transport through a single (non-interacting) electronic level.\cite{zimbovskaya2013transport,imry1999conductance,landauer1957spatial,jauho1994time} 

\subsection{Numerical example: single vibrational mode \label{onemode}}
\begin{figure*}
    \centering
    \includegraphics{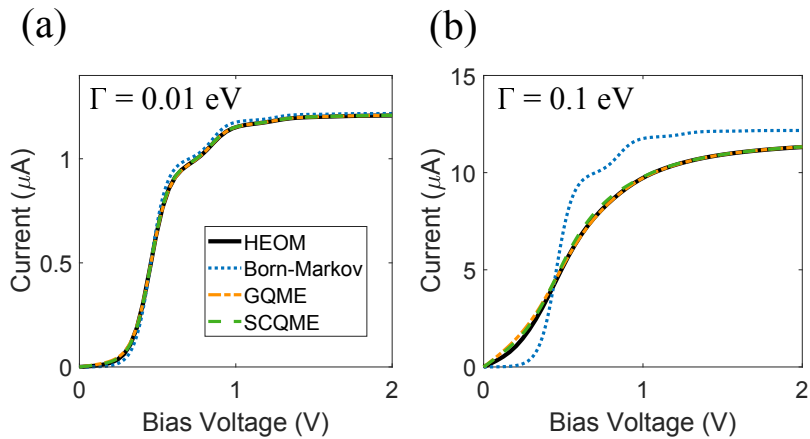}
    \caption{$IV$ characteristics calculated using the Born-Markov, GQME, SCQME and HEOM approaches for: $T= 300$ K, $\omega= 0.2$ eV, $g_0 = 0.12$ eV, $\bar{\varepsilon}_0 = 0.228$ eV, and symmetric molecule-lead coupling $\Gamma_L = \Gamma_R = \Gamma$.\cite{schinabeck2016hierarchical} Number of Pad\'{e} exponents and the maximum limit in the sum of the vibrational exponent terms in the HEOM calculations: (a) $L_\mathrm{max} = 8$, $u_\mathrm{max} = 3$; (b) $L_\mathrm{max} = 9$, $u_\mathrm{max} = 3$.}
    \label{heom1}
\end{figure*}

In this section we study the performance of our self-consistent quantum master equation (SCQME) approach for modelling transport through molecular junctions. We will also compare its performance to the Born-Markov (BM) quantum master equation (where $\Phi = 0$), the generalised quantum master equation (GQME) from Part I of this work (where $\Phi = \Gamma$), and the numerically-exact HEOM calculations (in which, for consistency, we also assume a thermalised vibrational environment).

For simplicity, we consider the case of coupling to a single molecular vibrational mode, i.e.~we set 
\begin{equation}
    \mathcal{J}(\omega) = \lvert g_0\rvert^2 \delta(\omega - \omega_0)~,
\end{equation}
where $\omega_0$ is the frequency and $g_0$ the coupling strength of the mode in question, and apply the bias voltage symmetrically: $\mu_l = \pm \lvert e \rvert V_\mathrm{b}/2$.

We first consider the $IV$ characteristics obtained using the aforementioned approaches.
We begin by considering the case of weak vibrational coupling ($g_0/\omega_0 = 0.6$). 
As shown in Fig.~\ref{heom1}, the GQME and SCQME approaches predict very similar behaviour which closely overlaps with that predicted by the HEOM approach. On the other hand, and as expected, the Born-Markov approach fails to accurately capture the current-voltage characteristics, especially in the case of stronger molecule-lead coupling, i.e.~when $\Gamma \gg k_\mathrm{B}T$, Fig.~\ref{heom1}(b).
Given that the GQME and SCQME approaches both yield the exact (Landauer-B\"uttiker) result in the absence of vibrational coupling, their very good agreement with the HEOM calculations (for relatively small $g_0/\omega_0$) is hardly surprising.
\begin{figure*}
    \centering
    \includegraphics{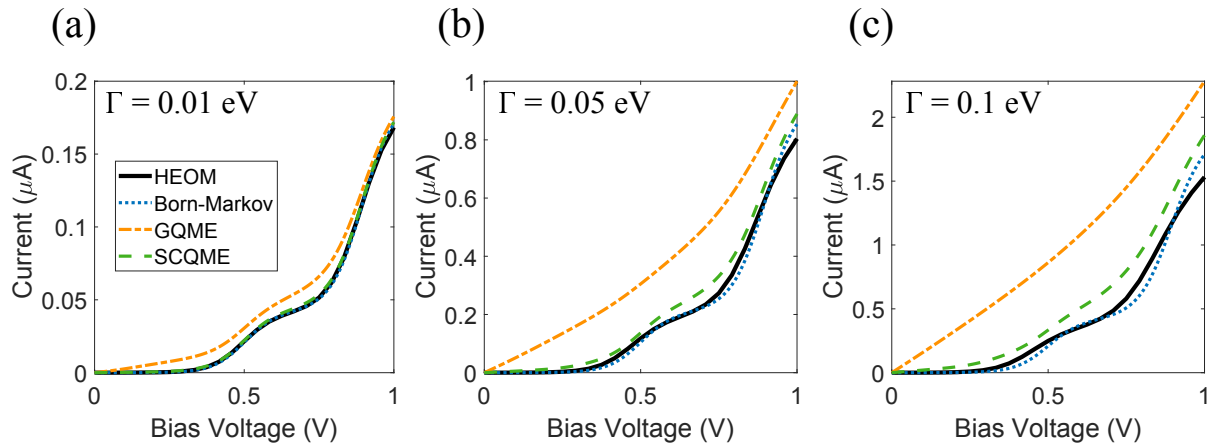}
    \caption{$IV$ characteristics calculated using the Born-Markov, GQME, SCQME and HEOM approaches for: $T= 300$ K, $\omega= 0.2$ eV, $g_0 = 0.4$ eV, $\bar{\varepsilon}_0 = 0.25$ eV, and symmetric molecule-lead coupling $\Gamma_L = \Gamma_R = \Gamma$.\cite{schinabeck2016hierarchical} Number of Pad\'{e} exponents and the maximum limit in the sum of the vibrational exponent terms in the HEOM calculations: (a) $L_\mathrm{max} = 8$, $u_\mathrm{max} = 10$; (b) $L_\mathrm{max} = 10$, $u_\mathrm{max} = 10$; (c) $L_\mathrm{max} = 9$, $u_\mathrm{max} = 10$.}
    \label{2heom}
\end{figure*}

We next to turn to the more challenging regime of strong-vibrational coupling ($g_0/\omega_0 = 2$). As shown in Fig.~\ref{2heom}, now the GQME approach very significantly overestimates the amount of lifetime broadening, see Section \ref{consider}.
In the case of  strong electron-vibrational coupling, therefore, the GQME approach fails to correctly describe the $IV$ characteristics [often rather spectacularly, as shown in Fig.~\ref{2heom}(c)].
Conversely, and perhaps somewhat unexpectedly, in this regime, the Born-Markov approach provides a much better approximation of the exact result. 
The SCQME theory clearly interpolates between the Born-Markov and GQME approaches, and in the case of strong electron-vibrational coupling delivers a result very close to that resulting from the Born-Markov approximation. Consequently, it appears to predict the shape of the $IV$ characteristics relatively well for small to moderate $\Gamma$.
The case of concurrently strong electron-vibrational ($g_0/\omega_0 >1$) and molecule-lead couplings ($\Gamma \gg k_\mathrm{B}T$) is considered in Fig.~\ref{2heom}(b) and (c). Unsurprisingly none of our perturbative approaches captures the charge transport behaviour in this regime.
\begin{figure*}
    \centering
    \includegraphics{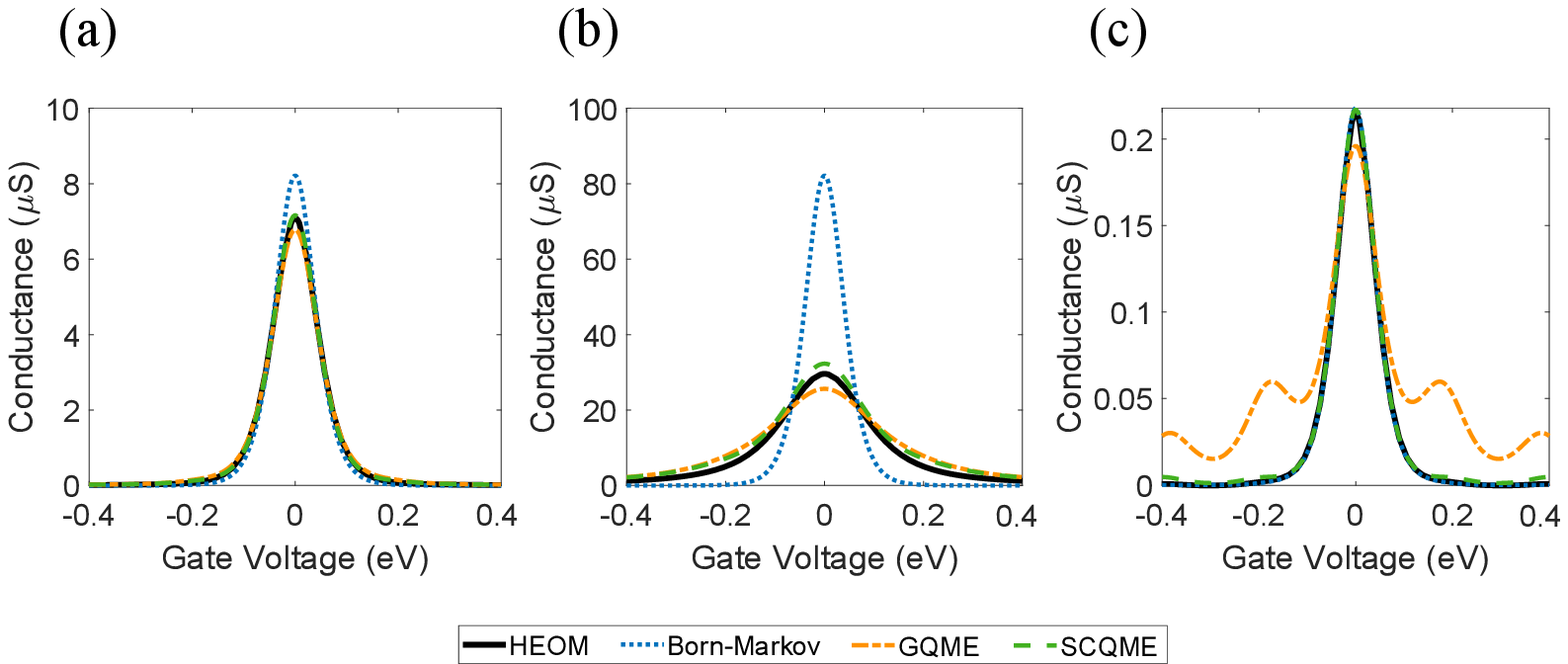}
    \caption{Zero-bias conductance traces calculated for: $T= 300$ K, $\omega= 0.2$ eV and symmetric molecule-lead coupling $\Gamma_L = \Gamma_R = \Gamma$. The position of the molecular level $\bar{\varepsilon}_0 = -\lvert e \rvert V_\mathrm{g}$. Other parameters: (a) $g_0 = 0.12$ eV, $\Gamma = 0.01$ eV;  (b) $g_0 = 0.12$ eV, $\Gamma = 0.1$ eV; (c)  $g_0 = 0.4$ eV, $\Gamma = 0.01$ eV. Number of Pad\'{e} exponents and the maximum limit in the sum of the vibrational exponent terms in the HEOM calculations: (a) $L_\mathrm{max} = 8$, $u_\mathrm{max} = 3$; (b) $L_\mathrm{max} = 9$, $u_\mathrm{max} = 3$; (c) $L_\mathrm{max} = 8$, $u_\mathrm{max} = 10$.}
    \label{gateheom}
\end{figure*}

Lastly, in Fig.~\ref{gateheom}, we consider the zero-bias conductance as a function of the gate voltage. These calculations largely support our earlier conclusions. The Born-Markov approach, due to the absence of lifetime broadening, overestimates the height of the conductance peak and underestimates the conductance away from resonance. This is especially pronounced for large $\Gamma$ and weak electron-vibrational coupling. Conversely, the GQME approach underestimates the conductance on resonance and overestimates it off resonance. Once again, it also fails in the case of strong vibrational coupling where it predicts spurious oscillations of the zero-bias conductance (which are clearly an artefact of this approach). The shortcomings of both the Born-Markov and GQME approaches are largely lifted by the self-consistent method developed in this work.

To summarise, the Born-Markov approach (which is perturbative in molecule-lead coupling) clearly fails for stronger molecule-lead coupling and generally significantly underestimates the current in the off-resonant regime. 
On the other hand, the GQME appears to be perturbative in $g_0/\omega_0$. It provides an accurate description of charge transport for small (or zero) $g_0/\omega_0$  but fails in the case of strong electron-vibrational coupling (where it typically overestimates the electric current, especially off resonance).
The self-consistent approach developed here interpolates between the above methods and consequently yields relatively accurate results across a much wider range of parameters.
We finish by noting that, unlike the hierarchical equations of motion technique, our theory trivially generalises to the case of multi-mode coupling (and indeed to the case of a continuous phononic spectral density) with little additional computational cost.

\subsection{General considerations \label{consider}}
Having demonstrated the effectiveness of our self-consistent approach (in the presence of electron-vibrational interactions) on a numerical example, let us discuss the properties of the self-consistent theory in more detail. 
The expression for the electric current derived in this work has a very similar structure to that derived in Part I. As stated before, the crucial difference between the two approaches is that here the amount of lifetime broadening depends on the strength of electron-vibrational interactions as well as the bias voltage and the position of the molecular energy level.
Physically, this accounts for the fact that the broadening associated with a given (electron-vibrational) transition should depend on the probability of this transition taking place.\cite{flensberg2003tunneling} For instance, one should expect the overall ground-state-to-ground-state transition to be associated with a much smaller degree of lifetime broadening in the case of strong electron-vibrational coupling than in the absence of electron-vibrational coupling (due to the poor Franck-Condon overlap between the states in the former case\cite{koch2005franck,koch2006theory}).

In the limit of high bias voltage or for the molecular energy level lying far away from the Fermi levels of the unbiased leads, it can be easily shown that $\mathrm{Re}[\nu_l] \rightarrow \Gamma_l/2$ and $\mathrm{Re}[\bar{\nu_l}] \rightarrow 0$ (or \textit{vice versa}). Then, the damping term, $\Phi$, in Eqs.~\eqref{R1} and \eqref{R2} becomes simply: $\mathrm{Re}\left[\Phi\right] \rightarrow \Gamma$.   
In these limits (i.e.~in the ``deep resonant" and the ``deep off-resonant" regimes), the self-consistent approach developed here reduces to the generalised approach derived in Part~I.

As stated above, the imaginary part of $\Phi$ induces a renormalisation of the position of the molecular energy level. Assuming the wide-band approximation, we have shown above that $\mathrm{Im}\left[\Phi\right]$ vanishes in the absence of electron-vibrational interactions.
For the case of coupling to individual (undamped) vibrational modes $\mathrm{Im}\left[\Phi\right]$ diverges (see Appendix \ref{appFou}). When $\mathrm{Im}\left[\Phi\right] $ converges, we find that it is on the order of $\Gamma$ and therefore (in the weak-coupling limit which is predominantly of interest here) constitutes a relatively small correction to $\bar{\varepsilon}_0$. This effect is therefore largely inconsequential to the present discussion.

Lastly, we turn to the limitations of our theory. First, as already briefly discussed, we have assumed that the vibrational environment is thermalised at all times. It is known (and has been demonstrated experimentally\cite{lau2015redox}) that non-equilibrium vibrational dynamics (especially of intra-molecular vibrational modes) can play an important role in steady-state transport through molecular junctions.\cite{hartle2011resonant} Nonetheless, our assumption is expected to be valid in the case of coupling to a solvent or substrate environment and when molecule-lead interactions are relatively weak.  
Second, our self-consistent approach was derived based on a (modified) second-order quantum master equation. 
It is thus expected to be valid in the regime of relatively weak molecule-lead coupling where individual electron transfers between the molecular structure and the metallic leads can be treated as non-adiabatic processes. 
This approach therefore becomes exact in the limit of weak molecule-lead coupling (i.e.~for $\Gamma\rightarrow 0$) when it coincides with the conventional second-order theory but also, as we have shown above, in the limit of vanishing electron-vibrational coupling. Conversely, one should indeed expect the self-consistent theory developed here to fail in the case of simultaneously  strong electron-vibrational coupling and considerable molecule-lead coupling.

\section{Marcus limit \label{highT}}
In Part I of this work, by considering the high-temperature limit, we have shown how lifetime broadening can be introduced into the conventional Marcus theory\cite{marcus1956theory,marcus1985electron,nitzan2006chemical} (which itself has been extensively used in the past to model the transport properties of molecular junctions).\cite{zhang2008single,migliore2011nonlinear,migliore2013irreversibility,thomas2019understanding,jia2016covalently,yuan2018transition,valianti2019charge}
In what follows, we demonstrate how lifetime broadening can be introduced into Marcus theory in a self-consistent fashion [and thus derive the self-consistent Marcus theory (scMarcus)]. We will then compare it to the conventional Marcus (also known as Marcus-Hush-Chidsey) theory and the generalised Marcus theory (gMarcus) derived in Part I of this work. Both of these alternative theories are for convenience given in Appendix \ref{appMarcus}.

\subsection{Derivation}
Similarly as we have done in the Part I of this work, we  take the high-temperature limit in the phononic correlation function $B(t)$. In particular, we approximate $\mathrm{coth}\left(\beta \omega/2\right) \approx 2/\beta\omega$, and assume a slowly-fluctuating environment: $\sin(\omega t) \approx \omega t$ and $\cos(\omega t) \approx 1 - \omega^2 t^2 /2$.\cite{may2008charge,taylor2018generalised} Then, the phononic correlation function becomes:
\begin{equation}
    B(t) \approx \exp(-\lambda t^2/\beta) \exp(-\mathrm{i}\lambda t) ~,
\end{equation}
where $\lambda$ is the Marcus reorganisation energy:
\begin{equation}
    \lambda = \int_0^\infty \mathrm{d}\omega \: \mathcal{J}(\omega)/\omega ~.
\end{equation}
Performing the one-sided Fourier transform, we obtain
\begin{multline} \label{PHII}
\Phi_\mathrm{M} = \sum_l \Gamma_l \left\{ \int_{-\infty}^\infty \dfrac{\mathrm{d}\epsilon}{2\pi} f_l(\epsilon) K_+(\epsilon)\ + \right. \\ \left. \int_{-\infty}^\infty \dfrac{\mathrm{d}\epsilon}{2\pi} [1 - f_l(\epsilon)] K_-^*(\epsilon)  \right\} ~,
\end{multline}
where $K_\pm(\epsilon)$ take the form,
\begin{multline}
    K_\pm(\epsilon) = \sqrt{\dfrac{\pi}{4\lambda k_{\mathrm{B}} T }} \exp\left( -\dfrac{[ (\epsilon - \bar{\varepsilon}_0) \mp \lambda]^2}{4\lambda k_{\mathrm{B}} T}\right)  \times \\ \left\{ 1 \pm \mathrm{i} \ \mathrm{erfi}\left( \dfrac{(\epsilon- \bar{\varepsilon}_0) \mp \lambda}{\sqrt{4\lambda k_\mathrm{B}T}}\right)\right\}~.
\end{multline}
In the above, $\mathrm{erfi}(x)$ denotes an imaginary error function [and the real part of $K(\epsilon)$ has the familiar Gaussian form]. Finally, the hopping rates become:
\begin{multline}
\gamma_l^\mathrm{M} =  2\ \mathrm{Re} \Bigg[ \Gamma_l \int_{-\infty}^\infty \dfrac{\mathrm{d}\epsilon}{2\pi} f_l(\epsilon)\times \\  \int_0^\infty \mathrm{d}\tau \ e^{+\mathrm{i}(\epsilon - \bar{\varepsilon}_0 -\lambda )\tau} e^{-\Phi^\mathrm{M} \tau} e^{-\lambda \tau^2/\beta} \Bigg]~,
\end{multline}
and similarly for $\bar{\gamma}_l^\mathrm{M}$.
Performing the one-sided Fourier transforms leads to:
\begin{align}
\begin{split}\label{Rr1}
 \gamma_l^\mathrm{M} =  2\ \Gamma_l \int_{-\infty}^\infty \dfrac{\mathrm{d}\epsilon}{2\pi} f_l(\epsilon)\ \mathrm{Re} \bigg[\sqrt{\dfrac{\pi}{4\lambda k_{\mathrm{B}} T}}\ \times \\ \exp\left( \dfrac{(\Phi_\mathrm{M} - \mathrm{i}\nu_+)^2}{4\lambda k_{\mathrm{B}} T}\right) \mathrm{erfc}\left(\dfrac{\Phi_\mathrm{M} - \mathrm{i}\nu_+}{\sqrt{4\lambda k_{\mathrm{B}} T}}\right)\bigg] ~;
\end{split}\\ 
\begin{split}
 \bar{\gamma}_l^\mathrm{M} = 2\ \Gamma_l \int_{-\infty}^\infty \dfrac{\mathrm{d}\epsilon}{2\pi} [1 - f_l(\epsilon)]\ \mathrm{Re} \bigg[\sqrt{\dfrac{\pi}{4\lambda k_{\mathrm{B}} T}} \ \times \\ \exp\left( \dfrac{(\Phi_\mathrm{M} - \mathrm{i}\nu_-)^2}{4\lambda k_{\mathrm{B}} T}\right) \mathrm{erfc}\left(\dfrac{\Phi_\mathrm{M} - \mathrm{i}\nu_-}{\sqrt{4\lambda k_{\mathrm{B}} T}}\right)\bigg] ~, \label{Rr2}
 \end{split}
\end{align}
where $ \nu_{\pm}=   \epsilon - \bar{\varepsilon}_0 \mp \lambda$ and $\mathrm{erfc}(x)$ is the complementary error function.
Similarly to the generalised Marcus theory, therefore, the molecular densities of states take the form of Voigt profiles, see Appendix \ref{appMarcus}. Now, however, the amount of Lorentzian-type broadening depends on the bias voltage and the position of the molecular level:
\begin{multline}
   \mathrm{Re}[\Phi_\mathrm{M}] = \sum_l \Gamma_l \bigg\{ \dfrac{1}{2} + \sqrt{\dfrac{\pi}{4\lambda k_{\mathrm{B}} T }} \int_{-\infty}^\infty \dfrac{\mathrm{d}\epsilon}{2\pi} f_l(\epsilon) \times  \\  \left[ \exp\left( -\dfrac{[ (\epsilon - \bar{\varepsilon}_0) - \lambda]^2}{4\lambda k_{\mathrm{B}} T}\right) - \exp\left( -\dfrac{[ (\epsilon - \bar{\varepsilon}_0) +
  \lambda]^2}{4\lambda k_{\mathrm{B}} T}\right)\right]  \bigg\} ~,
\end{multline}
In contrast to what has been discussed in Section \ref{onemode}, the renormalisations of the molecular energy level ($\mathrm{Im}[\Phi_\mathrm{M}]$) now converge within the wide-band approximation:
\begin{multline}
   \mathrm{Im}[\Phi_\mathrm{M}] =  \sum_l  \dfrac{\Gamma_l}{\sqrt{\lambda k_{\mathrm{B}} T }} \int_{-\infty}^\infty \dfrac{\mathrm{d}\epsilon}{2\pi} f_l(\epsilon) \times \\\left[ \mathcal{D}\left( \dfrac{(\epsilon- \bar{\varepsilon}_0) - \lambda}{\sqrt{4\lambda k_\mathrm{B}T}}\right)- \mathcal{D}\left( \dfrac{(\epsilon- \bar{\varepsilon}_0) + \lambda}{\sqrt{4\lambda k_\mathrm{B}T}}\right) \right] ~.
\end{multline}
where $\mathcal{D}(x)$ denotes the Dawson function.
In what follows we shall nonetheless disregard these renormalisations of the molecular energy level which are generally on the order of $\Gamma$ and are largely inconsequential for the subsequent considerations.

We note that in the limit of vanishing reorganisation energy $\lambda \rightarrow 0$, we once again recover the Landauer-B\"uttiker expression for the electric current (as demonstrated in Section \ref{Land}). On the other hand, if we were to ignore the self-consistent correction introduced here (i.e.~set $\Phi_\mathrm{M} = 0$) we would trivially recover the conventional Marcus theory. 
Similarly to the generalised Marcus theory, therefore, this (more sophisticated) self-consistent approach also unifies the Landauer-B\"uttiker and Marcus-type theories of molecular conduction.
\begin{figure*}
    \centering
    \includegraphics{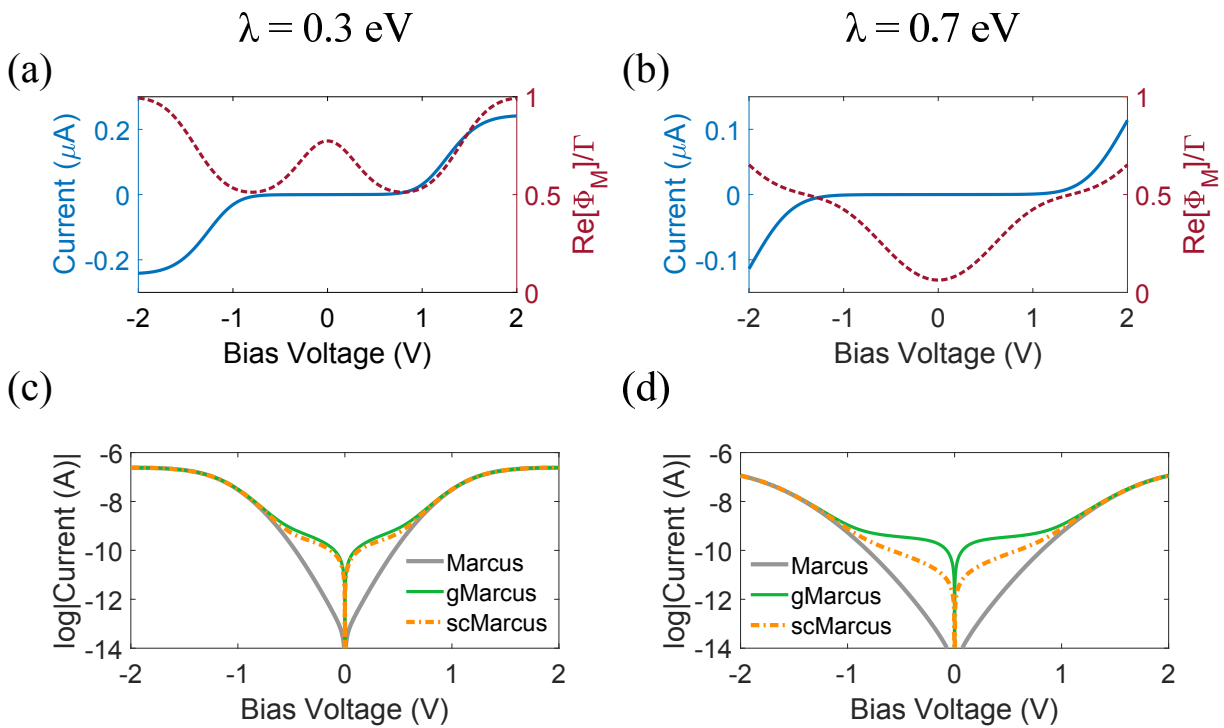}
    \caption{(a, b) $IV$ characteristics and the amount of lifetime broadening $\mathrm{Re}\left[\Phi_\mathrm{M}\right]/\Gamma$ as a function of bias voltage. (c,d) $IV$ characteristics on a logarithimc scale obtained using the self-consistent Marcus, generalised Marcus, and conventional Marcus theories. The reorganisation energy: (a,c) $\lambda = 0.3$ eV; (b,d) $\lambda = 0.7$ eV. Other parameters: $\bar{\varepsilon}_0 = 0.4$ eV, $\Gamma_\mathrm{L} = \Gamma_\mathrm{R} = 2$ meV, $T=300$ K.}
    \label{marcus1}
\end{figure*}

\subsection{Comparison with alternative approaches}
We proceed to compare the self-consistent approach developed here with the alternative Marcus-type theories given in Appendix \ref{appMarcus}.

First, using Eq.~\eqref{PHII} we can investigate when $\mathrm{Re}\left[\Phi_\mathrm{M}\right]\rightarrow\Gamma$, that is when the self-consistent theory developed in this work becomes identical to the generalised one developed in Part I. In particular, this is the case whenever $\mu_l \gg \bar{\varepsilon}_0 + \lambda$ or $\mu_l \ll \bar{\varepsilon}_0 - \lambda$ for both $l = \mathrm{L}, \mathrm{R}$. In other words, $\mathrm{Re}\left[\Phi_\mathrm{M}\right] = \Gamma$ in the deep off-resonant regime (i.e.~when $\mu_{\mathrm{L, R}} \gg \bar{\varepsilon}_0 + \lambda $ or when $\mu_{\mathrm{L, R}} \ll \bar{\varepsilon}_0 - \lambda $) and in the deep resonant regime (i.e.~when $\mu_{\mathrm{L}} \gg \bar{\varepsilon}_0 + \lambda $ and $\mu_{\mathrm{R}} \ll \bar{\varepsilon}_0 - \lambda $ or \textit{vice versa}). The biggest differences between the gMarcus and scMarcus approaches should therefore be present when the molecular energy level enters the bias window.
Furthermore, as discussed above, we note that the two theories become identical in the limit of vanishing reorganisation energy (where they both yield the Landauer-B\"uttiker expression). One should therefore also expect these two approaches to differ significantly in the case of relatively large $\lambda$.

In Figs.~\ref{marcus1}(a, b), we show the $IV$ characteristics calculated using the self-consistent Marcus approach for relatively weak molecule-lead coupling ($\Gamma\ll k_\mathrm{B}T$) and two different values of the reorganisation energy $\lambda$.
For simplicity, the bias voltage has been again applied symmetrically: $\mu_l = \pm \lvert e\rvert V_\mathrm{b}/2$.
In both cases we observe the usual behaviour: current suppression in the off-resonant regime which is lifted once the molecular level enters the bias window eventually followed by a saturation of the electric current at high bias.
Figs.~\ref{marcus1}(a, b) also show the amount of lifetime broadening ($\mathrm{Re}\left[\Phi_\mathrm{M}\right]/\Gamma$) as a function of the bias voltage. We can see that $\mathrm{Re}\left[\Phi_\mathrm{M}\right]\rightarrow\Gamma$ at large bias voltage (when the molecular energy level lies deep within the bias window).
At low bias voltage, the extent of lifetime broadening depends on the values of $\bar{\varepsilon}_0$ and $\lambda$, and is generally smaller for larger values of reorganisation energy, in accordance with our earlier discussion.

The comparison of the $IV$ behaviour predicted by the self-consistent, generalised and the conventional Marcus theory is shown (on a logarithmic scale) in Figs.~\ref{marcus1}(c, d).
For weaker environmental interactions ($\lambda = 0.3$ eV), the self-consistent Marcus theory will yield results similar to those of the generalised Marcus theory.
By contrast, in the case of significantly larger $\lambda$, the generalised Marcus theory significantly overestimates values of the off-resonant electric current, see discussion in Section \ref{onemode}.

We can also clearly see that, for both values of reorganisation energy, the conventional Marcus theory only correctly captures charge transport in the resonant regime (i.e.~for $V_\mathrm{b}>0.8$ eV). This should come as no surprise given that the conventional Marcus theory is only perturbative (to the second order) with respect to the molecular-lead interactions.

\begin{figure*}
    \centering
    \includegraphics{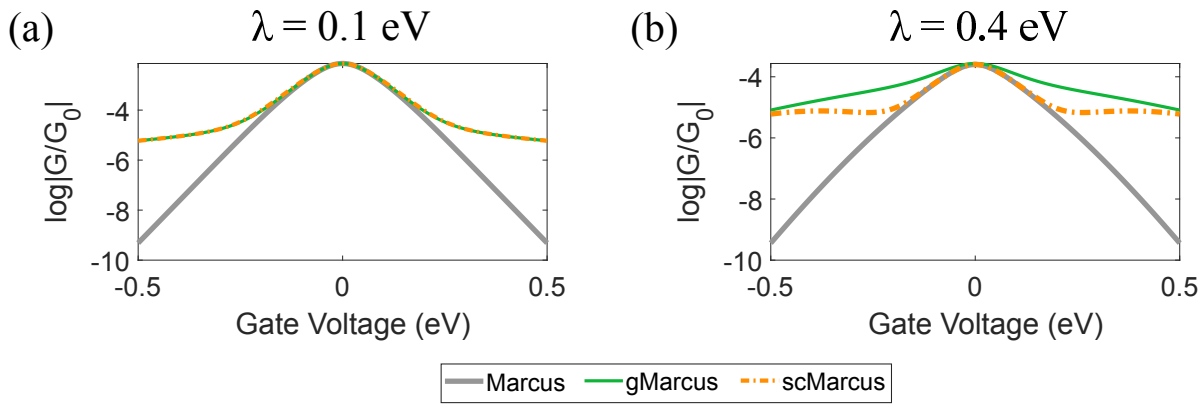}
    \caption{Zero-bias conductance plotted on a logarithimc scale obtained using the self-consistent Marcus, generalised Marcus, and conventional Marcus theories. The reorganisation energy: (a) $\lambda = 0.1$ eV; (b) $\lambda = 0.4$ eV. Other parameters: $\bar{\varepsilon}_0 = -\lvert e\rvert V_\mathrm{g}$, $\Gamma_\mathrm{L} = \Gamma_\mathrm{R} = 2$ meV, $T=300$ K.}
    \label{marcus2f}
\end{figure*}
It is also interesting to consider the zero-bias conductance predicted by the three approaches discussed here. Once again we consider the behaviour for two different values of the reorganisation energy.
As shown in Fig.~\ref{marcus2f}, for the relatively weak environmental coupling ($\lambda = 0.1$ eV) the generalised and self-consistent approaches are in an excellent agreement predicting virtually identical conductance values. Unsurprisingly, the conventional Marcus theory significantly underestimates the electric current off resonance although it is in a good agreement with the remaining approaches around the resonance.
Moving to the case of stronger environmental coupling ($\lambda = 0.4$ eV), we can see that the generalised Marcus theory significantly overestimates the width of the conductance peak. 
The self-consistent theory clearly interpolates between the conventional Marcus theory (with which it is in agreement close to resonance) and the generalised Marcus theory with which it agrees at large gate voltages (i.e.~far away from the resonance point).

\section{Conclusions \label{end}}
In this work, we have studied charge transport through molecular junctions modelled as a single site coupled to a set of vibrational modes. Assuming that the vibrational degrees of freedom remain thermalised at all times, we have derived a semi-analytical expression for the electric current in which lifetime broadening is introduced in a self-consistent fashion. 
This result improves on the theoretical framework developed in Part I of this work.
Considering the case of coupling to a single vibrational mode, we further compared our new self-consistent theory, the generalised theory from Part I of this work, and the conventional Born-Markov approach, to the results obtained using hierarchical equations of motion.
In doing so, we have demonstrated that the self-consistent theory significantly outperforms the considered alternative approaches at relatively  little additional (conceptual or computational) cost.

Similarly to we have done in Part I of this work, we have derived a modified version of the Marcus approach in which lifetime broadening is introduced in a self-consistent fashion. 
Our self-consistent theory (similarly to the generalised theory which was developed in Part I) also unifies the Marcus and Landauer-B\"uttiker descriptions of charge transport through molecular junctions (at least for the single-level model system considered here).

We have previously applied the theoretical framework from Part I of this work to study the phenomenon of thermoelectricity\cite{sowa2019marcus} as well as to model experimentally-observed charge transport through zinc porphyrin single-molecule transistors.\cite{thomas2019understanding}
Naturally, the self-consistent theory developed here could also be applied in these contexts.
Finally, it would be interesting to extend the type of theory developed in Part I and II of this work to longer (multi-site) molecular wires\cite{kilgour2015charge,sowa2017environment,kim2017controlling} (see also the very recent developments in Ref.~\onlinecite{liu2019generalized}), and to  account for vibrational non-equilibrium effects.\cite{lau2015redox,hartle2011resonant,sowa2018spiro,koch2006theory}

\begin{acknowledgments}
This work was supported by the National Science Foundation’s MRSEC program (DMR-1720319) at the Materials Research Center of Northwestern University. J.K.S.~also thanks EPSRC for the Doctoral Prize Award.
N.L.~acknowledges support from JST PRESTO, Grant no.~JPMJPR18GC,  and the RIKEN-AIST Joint Research Fund. E.M.G.~acknowledges funding from the Royal Society of Edinburgh and the Scottish Government.
\end{acknowledgments}

\appendix 

\section{Fourier transform of the phononic correlation function \label{appFou}}
In order to obtain the effective propagator, it is necessary to evaluate [in Eqs.~\eqref{rr1} and \eqref{rr2}] the Fourier transform of the phononic correlation function $B(\tau)$.

If the vibrational environment comprises a single (undamped) vibrational mode with frequencies $\omega_0$, i.e.
\begin{equation}
    \mathcal{J}(\omega) = \lvert g_0\rvert^2 \: \delta(\omega-\omega_0) ~,
\end{equation}
the relevant Fourier transform must be evaluated analytically. The correlation function can be written as:
\begin{equation}
    B_0(\tau)  =  \exp \Big\{ \alpha_0  \big[ \coth{\left(\beta\omega_0/2\right)} (\cos{\omega_0 t} - 1) - \mathrm{i}\sin{\omega_0 t} \big]  \Big\}~,
\end{equation}
where, for convenience, we have defined $\alpha_0 = ({g_0}/{\omega_0})^2$.
It will be convenient to follow Ref.~\onlinecite{mahan2013many} and write the single-mode correlation function as:
\begin{multline}\label{a3}
    B_0(\tau) = \exp\left[- \alpha_0   \coth{\left(\beta\omega_0/2\right)}\right] \times \\ \sum_{n = -\infty}^\infty I_{n} \left( \dfrac{\alpha_0 }{\sinh(\beta \omega_0/2)}  \right)  e^{n\beta\omega_0/2}  e^{-\mathrm{i}\tau n \omega_0}~,
\end{multline}
where $I_n$ is the modified Bessel function of the first type.

The one-sided Fourier transform of the single mode correlation function, $B_0$, is given by
\begin{multline}
\int_{0}^\infty \mathrm{d} \tau \: e^{\mathrm{i}(\epsilon-\bar{\varepsilon}_0)\tau}\: B_0(\tau) = \exp\left[- \alpha_0   \coth{\left(\beta\omega_0/2\right)}\right]\\ \times \sum_{n = -\infty}^\infty I_{n} \left( \dfrac{\alpha_0 }{\sinh(\beta \omega_0/2)}  \right) e^{n\beta\omega_0/2} \times \\\left\{ \pi \: \delta(\epsilon - \bar{\varepsilon}_0 - n \omega_0 ) +  \mathcal{P}\dfrac{\mathrm{i}}{\epsilon - \bar{\varepsilon}_0 - n \omega_0 }\right\}~.   
\end{multline}
Once inserted into Eqs.~\eqref{rr1} and \eqref{rr2}, the real part of $\Phi$ can be trivially found. 
On the other hand, the imaginary part diverges and will be disregarded henceforth.


\section{Alternative Marcus-type theories of transport \label{appMarcus}}
For convenience, here, we outline the conventional Marcus theory and its recently developed generalisation.
\subsection{Conventional Marcus approach}
The conventional Marcus (Marcus-Hush-Chidsey) theory describing electron transfer between a metallic lead and a molecular energy level (or \textit{vice versa}) can be obtained as the high-temperature limit of the second-order Born master equation (in the polaron-transformed frame), see Ref.~\onlinecite{sowa2018beyond} for a detailed discussion.
The electron hopping rates within this approach can be written as:
\begin{align}\label{rrr1}
\gamma_l &= 2\Gamma_l \int_{-\infty}^\infty \dfrac{\mathrm{d}\epsilon}{2\pi} f_l(\epsilon) K_+(\epsilon) ~;\\
\bar{\gamma}_l &= 2 \Gamma_l \int_{-\infty}^\infty \dfrac{\mathrm{d}\epsilon}{2\pi} [1 - f_l(\epsilon)] K_-(\epsilon)   ~,\label{rrr2}
\end{align}
where $K_\pm(\epsilon)$ has the familiar Marcus form:
\begin{equation}
    K_\pm(\epsilon) = \sqrt{\dfrac{\pi}{4\lambda k_{\mathrm{B}} T }} \exp\left( -\dfrac{[\lambda \mp (\epsilon - \bar{\varepsilon}_0)]^2}{4\lambda k_{\mathrm{B}} T}\right) ~.
\end{equation}

\subsection{Generalised Marcus theory}
In Part I of this work we have also demonstrated how lifetime broadening can be incorporated into the conventional Marcus theory. Within that (generalised Marcus) approach, the hopping rates have the form also given by Eqs.~\eqref{rrr1} and \eqref{rrr2} but with $K_\pm(\varepsilon)$ given by:\cite{sowa2018beyond,sowa2019marcus}
\begin{equation}
    K_\pm(\epsilon) = \mathrm{Re} \bigg[\sqrt{\dfrac{\pi}{4\lambda k_{\mathrm{B}} T}} \exp\left( \dfrac{(\Gamma - \mathrm{i}\nu_\pm)^2}{4\lambda k_{\mathrm{B}} T}\right) \mathrm{erfc}\left(\dfrac{\Gamma - \mathrm{i}\nu_\pm}{\sqrt{4\lambda k_{\mathrm{B}} T}}\right)\bigg]~, \label{gM}
\end{equation}
where $ \nu_{\pm}= \lambda \mp (\epsilon - \bar{\varepsilon}_0)$ and $\Gamma$ is again the lifetime broadening, $\Gamma = (\Gamma_\mathrm{L} + \Gamma_\mathrm{R})/2$.

%

\end{document}